\documentclass[12pt]{iopart}
\newcommand{\be}{\begin{equation}}
\newcommand{\ee}{\end{equation}}
\newcommand{\ba}{\begin{eqnarray}}
\newcommand{\ea}{\end{eqnarray}}

\usepackage[]{graphicx}

\begin{document}
\title[The motion of the Sun in interplanetary relativity experiments]{The effect
of the motion of the Sun on the light-time in interplanetary relativity
experiments}

\author{B Bertotti$^1$, N Ashby$^2$ and L Iess$^3$}
\address{$^1$ Dipartimento di Fisica Nucleare e Teorica,
Universit\`{a} di Pavia, via U. Bassi 6, 27100 Pavia, Italy}
\address{$^2$ Department of Physics, University of Colorado, Boulder, CO 80309-0390}
\address{$^3$ Dipartimento di Ingegneria Aerospaziale ed Astronautica,
Universit\`{a} La Sapienza, via Eudossiana 18, 00184 Rome, Italy}

\eads{\mailto{bb.142857@pv.infn.it}, \mailto{ashby@boulder.nist.gov},
\mailto{luciano.iess@uniroma1.it}}

\begin{abstract}
In 2002 a measurement of the effect of solar gravity upon the phase of coherent
microwave beams passing near the Sun has been carried out with the Cassini
mission, allowing a very accurate measurement of the PPN parameter $\gamma$.
The data have been analyzed with NASA's Orbit Determination Program (ODP) in
the Barycentric Celestial Reference System, in which the Sun moves around the
centre of mass of the solar system with a velocity $v_\odot$ of about 10 m/sec;
the question arises, what correction this implies for the predicted phase
shift. After a review of the way the ODP works, we set the problem in the
framework of Lorentz (and Galilean) transformations and evaluate the
correction; it is several orders of magnitude below our experimental accuracy.
We also discuss a recent paper \cite{kopeikin07}, which claims wrong and much
larger corrections, and clarify the reasons for the discrepancy.

\end{abstract}

\submitto{CQG}

\maketitle

\section{Introduction}

In 2002 the Cassini spacecraft, on its cruise to Saturn, has allowed an
outstanding measurement of the effect of solar gravity on the phase of a
coherent microwave beam sent from the ground antenna to the on-board
transponder and transmitted back. The data have been analysed with the Orbit
Determination Program (ODP), a numerical code developed at NASA's Jet
Propulsion Laboratory for accurate space navigation, including relativistic
effects; its algorithms are described in detail in \cite{moyer00} (referred to
here as the Manual). In the rest frame of the Sun the total phase change can be
measured by the light-time $t_2 - t_1$ between the events 1 and 2, the start
and the arrival of a photon in the up- or the down-link, respectively:

\be t_2 - t_1 = r_{12} + \Delta t = r_{12} + (\gamma +1) m \ln\frac{r_1 + r_2 +
r_{12}}{r_1 + r_2 - r_{12}} \label{eq:standard}. \ee Here $r_1$ and $r_2$ are
the distances of 1 and 2 from the Sun at their appropriate times $t_1$ and
$t_2$; $r_{12}$ is their Euclidian distance; $m = 1.43$ km is the gravitational
radius of the Sun\footnote{We mainly follow the notation of \cite{bblipt03} and
the Manual \cite{moyer00}; the velocity of light, however, is $c= 1$. Eq. (1)
of \cite{bblipt03} refers to the round-trip and has an additional factor 2; in
eq. (2) an obvious factor 2 is missing.}. This formula shows a characteristic
enhancement in the delay $\Delta t$ near conjunction, when the impact parameter
$b$ of the ray is much smaller than both $r_1$ and $r_2$; in this case the
argument of the logarithm in (\ref{eq:standard}) is about $4r_1r_2/b^2$, as in
eq. (1) of \cite{bblipt03}. This approximation, however, is not accurate enough
and is never used in the ODP. The fractional change in frequency $\Delta y$ is
essentially the rate of change of the light-time; as one can intuitively see,
and as fully discussed in \cite{bbgg92} (quoted as BG)\footnote{G. Giampieri
died on Sept. 2, 2006 at the age of 42. We take this opportunity to acknowledge
his outstanding and varied scientific contributions, in particular to BG.},
this frequency shift is of order $\alpha v$, where $\alpha$ is the deflection
angle and $v$ a linear combination of the velocities of the end points. The fit
of the Cassini data gave the result:

\be \gamma - 1 = (2.1 \pm 2.3)\times 10^{-5}. \label{eq:cassini}\ee An
important feature of the experiment was the use of a multi-frequency link,
which allowed an excellent elimination of the contribution of the plasma in the
solar corona. To date, this is by far the best measurement of the PPN parameter
$\gamma$, which has a crucial role in discriminating between alternative
theories of gravity.

Due to the other planets, in particular Jupiter, the Sun moves around the
centre of gravity of the solar system with a velocity $v_\odot \approx 5\times
10^{-8}$ (15 m/sec) and a time scale of several years. During a light-time
(about an hour) its motion is essentially uniform, so that its effect on the
light-time can be described by a Lorentz transformation and has nothing to do
with the intrinsic scattering dynamics. The short Nature paper \cite{bblipt03}
does not mention the motion of the Sun. The Manual does not address this
problem explicitly, nor evaluate the correction; however, we show that the ODP
does take into account this effect, and the fractional correction to the
one-way gravitational delay $\Delta t$ is of order $v_\odot$, quite below the
experimental sensitivity. The effect can also be described in terms of the
induced fractional frequency shift $\Delta y = \alpha v $, which attained the
value $3 \times 10^{-10}$ in the best passage. We show that the motion of the
Sun causes a change in $\Delta y$ of order $\alpha v v_\odot$, much smaller
than the frequency noise $\sigma_y \approx 10^{-14}$.  In a two-way experiment,
like Cassini's, even this small correction largely cancels out. In Sec. 4 we
discuss the recent paper \cite{kopeikin07}, which claims wrong and much larger
corrections, and point out the reasons for the discrepancy. Einstein's
prediction $\gamma =1$ is still unchallenged.

\section{Using the Orbit Determination Program}

The theoretical discussion in the main text of \cite{bblipt03} (eqs. (1) and
(2)) was given only to help physical understanding. The Block 5 receivers at
the ground station count the cycles of the phase of the received coherent beam
relative to the local frequency standard and determine the number of cycles in
successive time intervals. The duration of these intervals can be chosen from
0.1 sec to several hours (Sec. 13.3.1.3 of the Manual); during Cassini's
experiment it was 1 sec, but further averaging was done in the analysis. The
extraction of a time derivative from the actual phase count, which contains
fast-varying components, would be very delicate; the frequency is never
measured, nor is needed, since the mathematical expression of the predicted
light-time is at hand. This is why the ODP deals only with light-time. The
difference between Cassini's experiment and measurements of the radar delay
does not lie in a different gravitational observable, but in the fact that
Cassini uses a coherent radio beam and its phase as the main observable, while
planetary radar determines the arrival time of a wave packet through the peak
of its intensity. The strongest useful signal in Cassini's experiment occurred
on day 2 after conjunction, when the logarithm in (\ref{eq:standard}) was about
10, with a delay

$$ \Delta t = 1.43\,(\gamma + 1)\,10^6 \,\mathrm{cm};$$ thus the formal accuracy
$2.3\times 10^{-5}$ in $\gamma$ corresponds to an accuracy of 30 cm in $\Delta t$.

The ODP includes a very extensive and well tested orbit determination program,
with which the orbits $\mathbf{r}_i(t)$ of the centres of all relevant bodies
in the solar system, including the Sun and any spacecraft, are determined
numerically from previous and current observations. Following standard
astronomical usage and IAU recommendations, in particular the 2000 IAU
resolutions, this is carried out in the solar system Barycentric Celestial
Reference System (BCRS). The excellent paper \cite{soffel03} (referred here as
S) describes the procedure in detail. The expression S(8) for the metric tensor
is the basis for the definition of this frame; but, as T. Damour has discussed
in the first \cite{damour91} of four papers on relativistic frames, strictly
speaking the BCRS cannot be described in a single coordinate chart, but must
use a global chart for the dynamics of the gravitating bodies, and a local
chart for each gravitating body. In the linearized approximation we use,
however, this is not needed and S(8) has a global meaning. In the BCRS the
origin is at the centre of gravity C of the solar system; its axes are anchored
to the non rotating `celestial sphere' -- the International Celestial Reference
System (ICRS) -- as realized by very distant astronomical sources. Harmonic
coordinates are used: to the required order, gravity is described by a
(`small') spacetime tensor $h_{\mu\nu}$ in a fixed and conventional Minkowsky
background with metric $\eta_{\mu\nu} = \mathrm{diag}(-1,1,1,1)$. In this view
one can use Lorentz formalism; the word `null', the proper time $\tau$ and
raising and lowering spacetime indices refer to this metric. In addition, the
(cartesian) space components $h_{mn}$ are proportional to the unit matrix
$\delta_{mn}$ (the coordinates are `isotropic'). The time coordinate is the
Barycentric Coordinate Time (TCB), related to the Earth-bound Geocentric
Coordinate Time (TCG) by eq. S(58), in which, essentially, the Doppler shifts
due to the motion of the Earth and the gravitational shift due to external
gravitating bodies are corrected for.

In the present paper we only need to consider, besides the BCRS, the
heliocentric reference frame, neglecting quadratic terms in its gravitational
potential and the contributions of other masses. Then in isotropic coordinates
both $h_{00}$ and the space components $h_{mn}$ are functions only of the
distance $R$ from the origin. Note that in general relativity the field
equations do not uniquely determine the metric tensor in terms of the radial
coordinate; in our special coordinates this gauge freedom is forfeited. The
time coordinate $T$ is the privileged and invariant Killing time, with respect
to which the metric tensor is constant. The standard textbook expression for
the light-time (see, e.g., \cite{will93}) reads

\be T_2 - T_1 = R_{12} + \Delta T = R_{12} + (\gamma +1) m \ln\frac{R_1 + R_2 +
R_{12}}{R_1 + R_2 - R_{12}} \label{eq:standard_rest}. \ee The
coordinates\footnote{We use small letters with greek indexes from 0 to 3 to
denote spacetime quantities as geometrical objects in a generic coordinate
system; when time and space (boldface) components are written separately, we
use special coordinates: small latin letters in the barycentric frame and
capital letters in the rest frame of the Sun.} of the end events are $x_1^\mu
=(T_1, \mathbf{R}_1),\:x_2^\mu =(T_2, \mathbf{R}_2)$. It should be noted that
this formula holds only in isotropic coordinates.
\begin{figure}[ht]
\centering
\includegraphics*[width=5in]{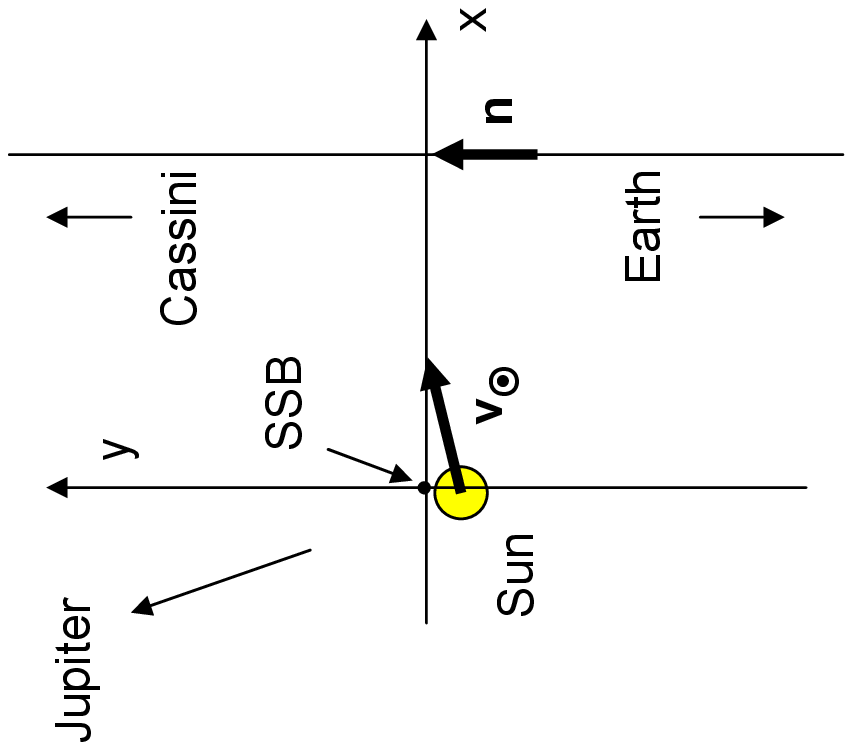}
\caption{The geometry of the experiment in the BCRS frame at epoch June 23-th
2002 20:00:00 UT, the first useful passage after conjunction in the Cassini
experiment. The center of the Sun moves in the ecliptic plane around the
barycenter of the solar system (SSB) in a roughly circular orbit with radius
$7.8 \times 10^8 $m; its velocity $v_\odot = 14.6 $ m/sec is approximately
orthogonal to the direction of Jupiter, and the orbital period is $\approx 12$
y.}
  \end{figure}

The ODP instead (see Fig. 1) computes the light-time in the BCRS, where the
centre of the Sun $\mathbf{r}_\odot(t)$ moves around C with a velocity $v_\odot
\approx 3\times 10^{-8}$; squares of $v_\odot$ can be neglected. As described
in Sec. 8.3.6 of the Manual, a special subroutine implements the formula
(\ref{eq:standard})\footnote{This corresponds to eq. (8-55) of the Manual,
which is the one actually coded in the program. Besides the Sun, (8-55) may
include the gravitational delays due to 10 other bodies. The argument of the
logarithm in (\ref{eq:standard}) includes a small correction of order $m$ which
is not relevant to our problem.} in 32 steps, an inclusion discussed at length
in \cite{hellings86}. The main object of the present paper is to evaluate the
difference in the light-time in the two frames.

In the Manual, Sec. 8.3.1.1, the three quantities $r_1$, $r_2$ and $r_{12}$ are
defined in barycentric coordinates as follows. The numerical code provides for
the bodies 1 and 2 and for the Sun the orbits $\mathbf{r}_1^C(t)$,
$\mathbf{r}_2^C(t)$, $\mathbf{r}_\odot^C(t)$ relative to C in the BCRS. The
simultaneous differences (eq. (8-62) of the Manual)

\be \mathbf{r}_1(t_1) = \mathbf{r}_1^C(t_1) - \mathbf{r}_\odot^C(t_1), \quad
\mathbf{r}_2(t_2) = \mathbf{r}_2^C(t_2) - \mathbf{r}_\odot^C(t_2)
\label{eq:ODP}\ee provide $r_1 =|\mathbf{r}_1(t_1)|$ and  $r_2
=|\mathbf{r}_2(t_2)|$; $r_{12}$ is the modulus of

\be \mathbf{r}_{12} = \mathbf{r}_2(t_2) - \mathbf{r}_1(t_1).
\label{eq:relative}\ee Note that differences between positions of bodies \emph{
taken at the same time} are invariant under Galilei transformations; hence
$r_1$, $r_2$ and $r_{12}$ have, up to $O(v_\odot^2)$, the same value in the
rest frame and the barycentric frame. We'll return to this important point
later.

It is useful to clarify the logical status of eq. (\ref{eq:standard}) and of
the subroutine used in the ODP to implement it, as explained in Sec. 8.3.2 of
the Manual. Given the arrival time $t_2$ at the point $\mathbf{r}_2(t_2)$, the
light-time solution must provide the starting time $t_1$, and the position and
velocity of the transmitter 1 at that time. A first approximation to $t_1$ is
given by the geometrical solution

$$ t_1 = t_2 - r_{12}(t_1), $$ where the argument of the distance indicates the
time at which 1 is taken; an error $\Delta t \approx 10^{-4}$ sec is made due
to the neglected gravitational delay. In this short time interval the velocity
of 1 is uniform and a linear correction  $\delta t_1$

$$ t_1 + \delta t_1 = t_2 - r_{12}(t_1) - \delta t_1 \frac{dr_{12}}{dt_1} -
\Delta t $$ is sufficient. The corrected value of the light-time is then
compared with the observations.

\section{Measuring the light-time in the barycentric frame}

We first comment about coordinates and geometrical objects. In the ODP scheme
-- the radial coordinate $r$ being fixed -- physical laws and statements are
invariant under the Lorentz group; to deal with this requirement we take the
geometric point of view, in which scalars, vectors and tensors -- like
$h_{\mu\nu}$ -- denote a geometrical object, independent of its coordinate
representation. Measured quantities are invariant scalars. The choice of
coordinates is conventional and free; for a vector, like $v_\mu = (v_0,
\textbf{v})$, they are indicated in round brackets, with the space components
in bold type.

We only need to deal with the spacetime vector $ \ell^\mu $ from 1 to 2, the
geometric object which summarizes the scattering dynamics of the experiment. In
the rest frame, where the Sun sits at the origin, its components are

\be \ell^\mu = x_2^\mu - x_1^\mu = (T_2-T_1,\mathbf{R}_{12}) = (R_{12}+ \Delta T,
\mathbf{R}_2 -\mathbf{R}_1) \label{eq:rest_expression}; \ee  if the mass $m$
vanishes

$$ \ell^\mu = (R_{12}, \mathbf{R}_{12}) $$ is null; gravity just adds a small,
time-like and positive contribution $(\Delta T, \mathbf{0})$ along the time
coordinate. An observer with (constant) spacetime velocity vector $v^\mu$
measures the light-time $- v_\mu \ell^\mu$; if the observer is at rest relative
to the Sun, $v^\mu_\odot = (1,\mathbf{0})$ and the previous result $T_2-T_1$ is
recovered; if it is at rest relative to the barycenter C, the ODP value
$t_2-t_1$ is obtained. Of course, the components of the light-time in the two
systems are related by a Lorentz transformation.

\cite{kopeikin07}, \cite{klioner03} and other papers take a different approach.
They apply the appropriate Lorentz transformation to the full trajectory of the
photon in the rest frame (as given, for instance, in \cite{will93}) and obtain
its expression in the barycentric frame; then they compute the elapsed time.
Delay and bending of the transformed orbit are then determined by the gravity
of a \emph{moving Sun}, with new gravitomagnetic terms appearing. This
complicates matters. A similar situation would arise in the classical
Rutherford scattering of an electron by a proton, normally described by the
Coulomb field of the latter, assumed at rest; of course, one would get the same
result in a frame where the proton moves, provided the magnetic field so
generated is taken into account. Just as magnetism is the direct consequence of
the fact that under Lorentz transformations the electromagnetic potential
behaves as a four-vector, so gravitomagnetic metric perturbations arise due to
Lorentz invariance and the tensorial character of the metric\footnote{For an
elementary derivation of gravitomagnetism from Schwarzschild linearized
solution, see \cite{bbpfdv03}, p. 571.}.

The analysis of Cassini's experiment has been carried out in the standard
framework of a Lorentz invariant theory. If Lorentz invariance is violated,
gravitomagnetism does not have the standard form; moreover, the problem of the
effect of the motion of the Sun on the gravitational delay must be addressed in
a way different from the one we follow below. KPSV suggest that, with its
excellent accuracy, Cassini's experiment may set better limits to such
violations. This would require a wise decision about the best theoretical
scheme and the appropriate parametrization; besides the PPN preferred frame
formalism, one has vector-tensor theories (see \cite{will93}) and
Lorentz-violating electromagnetism, on which there is a wide literature (see
\cite{ni77}, \cite{ni84}, \cite{ni05}, \cite{bailey04}). Current limits from
other experiments should also be taken into account. This program is outside
the present paper.

It is more convenient to skip the formal coordinate change in the trajectory
and to use just the light-time vector -- a geometrical object -- expressed as

$$ \ell^\mu = (t_2- t_1, \mathbf{x}_2 - \mathbf{x}_1) $$
 in the barycentric frame and as

$$ \ell^\mu = (T_2- T_1, \mathbf{R}_2 - \mathbf{R}_1) $$ in the rest frame of
the Sun. The barycentric components are related to (\ref{eq:rest_expression})
by a Lorentz transformation corresponding to the (small) velocity of the Sun
$\mathbf{v}_\odot$:

\ba t_2 - t_1  & = & T_2 - T_1 + \mathbf{v}_\odot\cdot \mathbf{R}_{12} +
O(v_\odot^2), \label{eq:time} \\
\mathbf{x}_2 - \mathbf{x}_1 &=& \mathbf{R}_{12}  + \mathbf{v}_\odot(T_2 - T_1)
+ O(v_\odot^2).\label{eq:space} \ea The space part is just a Galilei
transformation, arising from the fact that the end events occur at different
times. In Galileian relativity time is invariant; but this holds only if the
space component of a vector upon which the transformation operates is much
smaller than the time component. In our case $\ell^\mu$ is almost null and
(\ref{eq:time}) must be taken into account. The transformation conserves proper
length to the appropriate approximation:

\be  (t_2 - t_1)^2 -|\mathbf{x}_2 -\mathbf{x}_1|^2 = (T_2 - T_1)^2 -R_{12}^2
+O(v_\odot^2);\ee or, neglecting squares of the delay,

\be \Delta t\,|\mathbf{x}_2 - \mathbf{x}_1| = \Delta T R_{12}. \label{eq:diff}
\ee Now

$$\mathbf{x}_2 - \mathbf{x}_1 = \mathbf{r}_2^C(t_2) - \mathbf{r}_1^C(t_1),$$
the space component of $\ell^\mu$ in barycentric coordinates, obviously is not
Galilei invariant; but the Manual shrewdly uses $\mathbf{r}_{12}$ instead,
which, being built with simultaneous vectors, almost compensates the Galilei
transformation (\ref{eq:space}). Indeed, from (\ref{eq:relative})

\be \mathbf{r}_{12} = \mathbf{x}_2 - \mathbf{x}_1 - \mathbf{v}_\odot(t_2 - t_1); \ee
comparing it with (\ref{eq:space}), we see that
$$\mathbf{r}_{12}- \mathbf{R}_{12} = O(v_\odot^2).$$
Let $\mathbf{n} = \mathbf{r}_{12}/r_{12}$ be the (Galilei invariant!) unit
vector along the unpertubed ray from 1 to 2, as computed in the Manual. Eq.
(\ref{eq:space}) gives, to $O(v_\odot^2)$,

$$ |\mathbf{x}_2-\mathbf{x}_1| =R_{12} + \mathbf{v}_\odot \cdot \mathbf{n}
(T_2-T_1),$$ so that

\be \Delta t = \Delta T (1 - \mathbf{v}_\odot \cdot \mathbf{n}),\label{eq:true}
\ee which is our final result. It corresponds to a negligible change of about a
millimiter in light-time. On p. 8-28 the Manual acknowledges the neglect of
this correction: \emph{The error in the calculated delay due to ignoring the
barycentric velocity $v_\odot$ of the gravitating body has an order of
magnitude equal to the calculated delay multiplied by the velocity of the
body/$c$}. Our procedure clearly shows the gist of the problem, how the delay
transforms under a Lorentz transformation with a small velocity. As C. M. Will
has pointed out to us, a simpler proof is based upon his eq. (7.3) in
\cite{will93}, which says how the perturbation in the coordinate of the photon
along the ray changes with time; in the barycentric frame (7.3) acquires an
additional term due to the time-space components of the metric and proportional
to $m/r$; this leads to our result. Note also that in a two-way experiment,
like Cassini's, in the down-link $\mathbf{n}$ is almost equal and opposite its
up-link value and the two contributions essentially cancel out; a net result
arises from second order terms $O(v_\odot^2)$ in the Lorentz transformation.

Eq. (\ref{eq:true}) does not say, however, how to evaluate the delay in terms
of the distances $r_1, r_2$ and $r_{12}$ obtained in the BCRS. We need their
expressions in a generic frame, such that they reduce to $R_1$, $R_2$ and
$R_{12}$ in the rest frame. In the present context, invariance under the
approximate Lorentz group (\ref{eq:time}) and (\ref{eq:space}) is sufficient.
Then the ODP expressions (\ref{eq:ODP}) and (\ref{eq:relative}) provide the
straightforward answer: in fact they are constructed with simultaneous
differences and (\ref{eq:space}) reduces to Galilei transformations.

It interesting to discuss the fully invariant case. Distances arise here
because the Sun affects a photon essentially through the scalar potential $U$,
the fundamental solution of D'Alembert's equation. In the rest frame and, say,
for a photon at the start event $x_1^\mu = (T_1, \mathbf{R}_1)$,

\be U = \frac{m}{|\mathbf{R}_1|} =\frac{m}{R_1}. \ee We want to express it in a
generic frame, with the Sun moving with an arbitrary (but constant)
four-velocity $v_\odot^\mu = dX^\mu/d\tau$. The cartesian distance $R_1$ must
be replaced with \emph{the unique} quantity which (i) is Lorentz invariant and
(ii) reduces to $R_1$ when $v_{\odot \mu} =(-1, \mathbf{0})$ (the minus sign
being required by the metric $\mathrm{diag}(-1,1,1,1)$). This is a textbook
problem in electrodynamics (e. g., see \cite{landau75}, Sec. 63). Let $X^\mu_1$
be the (ante-dated) intersection of the (straight) world-line of the Sun with
the past light-cone of the start event $x_1^\mu = X_1^\mu + \ell_1^\mu$;
$\ell_1^\mu$ is the corresponding, future-pointing null vector. The required
`distance' is the invariant

\be r_1^\star =\left\vert v_{\odot \mu}(x_1^\mu - X^\mu_1)\right\vert =\left\vert
v_{\odot \mu}\ell_1^\mu \right\vert\label{eq:distance}. \ee Indeed, since

$$ x_1^0 - X^0_1 = |\mathbf{x}_1 - \mathbf{X}_1|  $$ (elapsed
time is equal to distance), it obviously fulfils (ii). The future light-cone
would give the same result.

In the appropriate slow motion approximation $v_{\odot \mu} = (-1,
\mathbf{v}_\odot)$ and (\ref{eq:distance}) reads

\be r_1^\star = x_1^0 - X_1^0 - \mathbf{v}_\odot \cdot (\mathbf{x}_1 -
\mathbf{X}_1) + O(v_\odot)^2. \ee $ x_1^0 - X_1^0$, the antedated distance
between the event 1 and the Sun, differs from the simultaneous value by an
amount equal to the distance traveled by the Sun in a time equal to the
distance itself; hence $r_1^\star$ is, to $O(v_\odot)$, just the simultaneous
distance $r_1$ (see (\ref{eq:ODP}) used in the ODP. In other words, the
first-order correction to Lienard-Wiechert potential due to a slowly moving
source vanishes\footnote{For the full expansion of the retarded potential in
powers of $1/c$, in which the term $O(1/c)$ is missing, see \cite{eddington60},
supplementary note 10 in the Appendix.}.

The fully invariant form of $r_{12} = |\mathbf{r}_2 - \mathbf{r}_1|$ is

\be r_{12}^\star = |v_{\odot \mu}\ell^\mu| = |v_{\odot \mu}(x_2^\mu - x_1^\mu)|
\label{eq:full}; \ee indeed, in the rest frame $x_1^\mu =(T_1, \mathbf{R}_1)$ and
$x_1^\mu =(T_2, \mathbf{R}_2)$, with $T_2- T_1 = |\mathbf{R}_2 - \mathbf{R}_1| =
R_{12}$. We now show that in the slow motion approximation

\be r_{12}^\star = x_2^0 - x_1^0 - \mathbf{v}_\odot \cdot (\mathbf{x}_2-
\mathbf{x}_1) + O(v_\odot)^2 \label{eq:12} \ee reduces to modulus of the ODP
(\ref{eq:relative}). We must use the simultaneous relative vector $\mathbf{r}_1
= \mathbf{x}_1 - \mathbf{X}(x_1^0)$, which differs from the space part of
$\ell_1^\mu$

$$ \mbox{\boldmath $\ell_1$} = \mathbf{x}_1 - \mathbf{X}_1 =\mathbf{x}_1 -
\mathbf{X}(x_1^0 - r_1) = \mathbf{r}_1 + r_1\mathbf{v}_\odot $$ because of the
ante-dated time argument in the position $\mathbf{X}(x^0)$ of the Sun. With a
similar first-order expansion we get

$$ \mathbf{X}_2 - \mathbf{X}_1 = \mathbf{X}(x_2^0- r_2) - \mathbf{X}(x_1^0- r_1)
= ( x_2^0 - x_1^0 - r_2 + r_1)\mathbf{v}_\odot = (r_{12} - r_2 +
r_1)\mathbf{v}_\odot, $$ where we have used

$$ \mathbf{X}(x^0_2) - \mathbf{X}(x^0_1) = \mathbf{v}_\odot (x_2^0- x_1^0), $$
$x_2^0 - x_1^0 = |\mathbf{x}_2 - \mathbf{x}_1|$ and the fact that $\mathbf{x}_2
- \mathbf{x}_1$ differs from $\mathbf{r}_2 - \mathbf{r}_1 = \mathbf{r}_{12}$ by
$O(v_\odot)^2$.
Then

\ba r_{12}^\star &=&|\mathbf{x}_2 - \mathbf{x}_1| - \mathbf{r}_{12} \cdot
\mathbf{v}_\odot = |\mathbf{r}_{12} + \mathbf{X}_2 - \mathbf{X}_1 +
(r_2-r_1)\mathbf{v}_\odot| - \mathbf{r}_{12} \cdot \mathbf{v}_\odot = \nonumber
\\ &= &|\mathbf{r}_{12} + r_{12}\mathbf{v}_\odot| - \mathbf{r}_{12} \cdot
\mathbf{v}_\odot = r_{12} + O(v_\odot)^2. \nonumber \ea We see that the ODP
variable $r_{12}$ is the appropriate approximation of a fully invariant
quantity and describes correctly the gravitational interaction.

\section{About Galilei invariance }

KPSV make two different claims about the effect of the motion of the Sun. Their
expression of the gravitational delay in the barycentric frame KPSV(25)

\be \Delta_{_{K}} t = (\gamma +1) m \ln\frac{r_1 + r_2 + r - \mathbf{r}\cdot
\mathbf{v}_\odot}{r_1 + r_2 - r +\mathbf{r}\cdot \mathbf{v}_\odot}.
\label{eq:kopeikin} \ee  differs from our analysis in two ways. First, it is
based upon the coordinate transformation KPSV(11), which does not provide the
required, first order difference between $t$ and $T$ (eq. (\ref{eq:time}));
this makes it impossible to get the true correction (\ref{eq:true}). More
important, the quantity $r_{12}$ used in the ODP is not their $r$, which, in
our notation, is $|\mathbf{x}_2 - \mathbf{x}_1|$. This quantity is not a
Galilei invariant and cannot appear in the argument of the logarithm. This
argument must depend on the coordinates of the photon \emph{relative} to the
Sun. Let us see the consequences of this claim. In the extreme conjunction case
($b \ll (r_1,r_2))$ the correction in the denominator of the argument of the
logarithm prevails and (\ref{eq:kopeikin}) reads

$$ \Delta_{_{K}} t = (\gamma +1)m \left(\ln\frac{4r_1r_2}{b^2}- \mathbf{v}_\odot
\cdot \mathbf{n}\frac{r_1r_2}{b^2}\right). $$ The claimed correction $\approx m
v_\odot$ is similar to the true correction (\ref{eq:true}), but is enhanced
when $b$ is small.

In their second claim, KPSV consider the description of the experiment in terms
of the effect that the deflection has on the frequencies recorded by two very
distant clocks (see \cite{bbgg92}). Of course the deflection angle

\be \alpha = 2(\gamma + 1)\frac{m}{b} \ee is meaningful only in the extreme
conjunction approximation $b \ll (r_1, r_2)$, which we assume here for the sake
of illustration. This gives also, in order of magnitude, the change in the
angle between the velocity $\mathbf{V}$ of a far-away clock \emph{relative to
the Sun} and the beam, and corresponds to a (one-way) fractional frequency
change (for a grazing ray and the Earth's orbital velocity) of order

\be \Delta y \approx \alpha V \approx 8 \times 10^{-10}.\ee Explicitly (see eq. (22) of
BG), in the rest frame we have

\be \Delta y = \frac{(\mathbf{V}_2\cdot \hat{\mathbf{b}}) r_1 -
(\mathbf{V}_1\cdot \hat{\mathbf{b}}) r_2}{r_{12}}\:\frac{2(\gamma +1)m}{b}.
\label{eq:BG} \ee In the following, for simplicity, we only consider frequency
measurements in terms of the proper times of 1 and 2. Now $\Delta y$, a ratio
of proper frequencies, is a Lorentz-invariant scalar; to get its expression in
a slowly moving frame, note first that the time coordinate is not involved, so
that only Galilei transformations need to be considered. We just need \emph{the
unique} quantity which (i) is Galilei invariant and (ii) reduces to
(\ref{eq:BG}) in the rest frame. The impact parameter vector is the component
of $\mathbf{r}_2$ (or, equivalently, $-\mathbf{r}_1$) along the ray:

\be \mathbf{b} = \mathbf{r}_2 - (\mathbf{n}\cdot \mathbf{r}_2) \mathbf{n}.\ee This is
invariant. Secondly, and trivially, the rest-frame vectors $\mathbf{V}_1$ and
$\mathbf{V}_2$ should be replaced with their invariant forms

\be \mathbf{v}_1 - \mathbf{v}_\odot =\frac{d\mathbf{r}_1(t_1)}{dt}, \quad \mathbf{v}_2
- \mathbf{v}_\odot =\frac{d\mathbf{r}_2(t_2)}{dt}. \ee

KPSV claim that in the barycentric frame, due to the motion of the Sun, there
is an additional correction

\be \Delta_K y \approx \alpha v_\odot \approx 3 \times 10^{-13}.
\label{eq:claim}\ee The fact that the impact parameter vector (parallel to the
deflection vector $\mbox{\boldmath $\alpha$}$) is a Galilei invariant seems to
be ignored; it is defined in the rest frame in terms of absolute coordinates,
with no reference to the Sun. Then in the barycentric frame the correction
KPSV(18) arises, which produces a small correction $O(v_\odot)$ in the
frequency shift. Moreover, absolute velocities are used in the analogue of
(\ref{eq:BG}), which, after a Galilei transformation, produces KPSV(21), the
basis for their main claim. Of course, the claimed correction (\ref{eq:claim})
is absorbed in (\ref{eq:BG}) using the \emph{relative} velocities $\mathbf{v}_1
- \mathbf{v}_\odot, \mathbf{v}_2 - \mathbf{v}_\odot$.

In Cassini's experiment the standard deviation of the residual frequency noise
(at an integration time of 1000 sec; see Fig. S2 of the supplementary
material\footnote{http://www.nature.com
/nature/journal/v425/n6956/suppinfo/nature01997.html.} of \cite{bblipt03}) was
$\sigma_y= 0.7 \times 10^{-14}$. While the magnitude of the correction
(\ref{eq:claim}) stated by KPSV is 40 times larger, the paper does not even
suggest a violation of general relativity. In the experiment $N =1094$ data
points, obtained after several pre-processing steps, have been used, with a
remarkably Gaussian distribution of the frequency residuals, as impressively
shown by Fig. 3 of \cite{bblipt03}. The sensitivity of $\gamma$ from each data
point is difficult to ascertain, but surely the final error in $\gamma$ which
results from the fit is \emph{smaller} than $\sigma_y$. KPSV, instead, state
the opposite (top of p. 280) and thereby avoid dangerous conclusions. In our
view, if the ODP dynamical model were as inaccurate as claimed by KPSV, the
inescapable conclusion would be that the good orbital fit and the agreement
with the standard prediction are the result of an exceedingly unlikely chance;
accepting the KPSV correction would imply the first experimental violation of
general relativity.

\section{Conclusion}

Cassini's experiment, by far the most accurate attempt to measure the PPN
parameters, did not pose a threat to Einstein's theory of gravitation, nor
answer the major question, at what level and how it is violated. The
measurement of the deflection parameter $\gamma$ has a crucial importance;
although no theoretical prediction is at hand about its deviation from unity,
alternative theories based upon a long range scalar field require $\gamma <1$.
In the future steady improvements in instrumentation, in particular space
astrometry and the use of optical links in interplanetary space, will allow
much more accurate measurements, and new relativistic space missions are under
construction or in the planning phase; see \cite{turyshev08} and \cite{ni08}
for reviews. We quote, in particular, the astrometric mission GAIA
\cite{mignard01}, the experimental program \emph{MORE} \cite{milani02} in the
\emph{BepiColombo} mission to Mercury and the project \emph{LATOR}
\cite{turyshev04}. Optical interferometric measurements, similar to those under
development for gravitational wave detection (the LISA project), are planned
for the ASTROD mission (\cite{ni04}, \cite{ni06}, \cite{ni07}). Binary pulsars,
whose gravitational field is much stronger, will also play a role in testing
gravitational theories.

A numerical code for relativistic orbit determination more accurate than the
ODP is surely needed. Aside from the correction (\ref{eq:true}) in the
light-time, terms in the metric quadratic in the gravitational radius $m$ must
be included. The second order gravitational deflection depends on the radial
gauge (see \cite{bodenner03}); in the isotropic gauge, for a grazing ray, it
has the value $\alpha_{\mathrm{quad}} = 1.2 \times 10^{-11}$, corresponding to
a frequency shift $\approx v_\odot \alpha_{\mathrm{quad}} = 10^{-15}$, an
accuracy easily accessible with laboratory standards. We should also mention
that in conjunction experiments, besides the obvious smallness parameter
$m/R_\odot = 2\times 10^{-6}$, the smallness of $b/r$ may produce an
enhancement in the light-time. At the first order in $m$ this is already
apparent in the logarithmic term of (\ref{eq:standard}); but there is also a
quadratic correction to the light-time of order $mr/b^2$, which may produce an
effect in Cassini's experiment about the same as the formal error. A correction
of this kind is included in the ODP. Finally, the complex procedures used in
the definition and the construction of the BCRS and the TCB need to be
revisited and improved both at the fundamental level and in the numerical
implementation.

\section*{Acknowledgements} We are grateful to S. Klioner for extensive
discussions, in particular about his paper \cite{klioner03}; to C. M. Will and
P. Bender for their comments. The work of B. Bertotti and L. Iess has been
supported in part by the Italian Space Agency (ASI).

 \end{document}